\documentclass[prl,twocolumn,superscriptaddress,showpacs]{revtex4}
\usepackage{graphicx} \usepackage{amsmath} \usepackage{color} \usepackage{array}
\usepackage{color}

\renewcommand\k{\mathbf{k}}

\begin{document}

\title{Glassy dynamics in geometrically frustrated Coulomb liquids without
disorder}

 \author{Samiyeh Mahmoudian} \affiliation{Department of Physics and National
 High Magnetic Field Laboratory, Florida State University, Tallahassee, Florida
 32306, USA} 
 \author{Louk Rademaker} \affiliation{Kavli Institute for
 Theoretical Physics, University of California Santa Barbara, CA 93106, USA}
 \author{Arnaud Ralko} \affiliation{Institut N\'{e}el-CNRS and Universit\'{e}
 Joseph Fourier, Bo\^{i}te Postale 166, F-38042 Grenoble Cedex 9, France}
 \author{Simone Fratini} \affiliation{Institut N\'{e}el-CNRS and Universit\'{e}
 Joseph Fourier, Bo\^{i}te Postale 166, F-38042 Grenoble Cedex 9, France}
 \author{Vladimir Dobrosavljevi\'{c}} \affiliation{Department of Physics and
 National High Magnetic Field Laboratory, Florida State University, Tallahassee,
 Florida 32306, USA}

\begin{abstract} We show that introducing long-range Coulomb interactions
immediately lifts the massive ground state degeneracy induced by geometric
frustration for electrons on quarter-filled triangular lattices in the
classical limit. Important consequences include the stabilization of a
stripe-ordered crystalline (global) ground state, but also the emergence of very
many low-lying metastable states with amorphous ``stripe-glass" spatial
structures. Melting of the stripe order thus leads to 
a frustrated Coulomb liquid at intermediate temperatures, showing
remarkably slow (viscous) dynamics, with very long relaxation times growing in
Arrhenius fashion upon cooling, as typical of strong glass formers. On
shorter time scales, the system falls out of equilibrium and displays the
aging phenomena characteristic of supercooled liquids above the glass
transition. Our results
show remarkable similarity with the recent observations of charge-glass
behavior in ultra-clean triangular organic materials of the  $\theta$-(BEDT-TTF)$_2$ 
family. \end{abstract}


\maketitle

Metastability, slow relaxation, and other features of
glassy dynamics are often observed in electronic systems at the brink of the
metal-insulator transition \cite{dobrosavljevic2012conductor}. These effects,
however, are typically attributed to disorder caused by impurities or defects,
rather than being an intrinsic feature of strongly interacting electrons.
Indeed, ``Coulomb glass" behavior \cite{pastor99,Grempel} is well established in
disordered insulators \cite{ovadyahu2000prl,popovic2002prl}; in other cases
metastability can be caused by disorder-dominated phase separation in presence
of competing orders \cite{dagotto-2005}.

A more intriguing possibility was proposed in the heyday of cuprate superconductivity, with the idea of
``Coulomb-frustrated phase separation"  in lightly doped Mott insulators
\cite{sokol1987phase,emery1993physica}. It suggested the possibility of
spontaneous (disorder-unrelated)  formation of many complicated patterns of
charge density, such as bubbles and stripe crystals \cite{spivak-prl05}, or even
stripe glasses \cite{schmalian-prl00}. Unfortunately, no conclusive theoretical or
experimental evidence emerged to support the existence of such phase separation,
which long remained more of a theorist's dream than an accepted mechanism for
metastability in electronic systems.

Glassy freezing without disorder, on the other hand, is well established in
several systems with geometric frustration, most notably the supercooled liquids
\cite{Debenedetti2001}. A natural question thus emerges: can sufficient
geometric frustration cause disorder-free glassy behavior of electrons, in
(many) situations where phase separation effects are not relevant? Geometric
frustration arises, for example, in spin systems on triangular lattices
\cite{Balents2010,Ramirez1994}, sometimes leading to exotic phases like spin
liquids \cite{Shimizu2003}; related frustration-driven phenomena in the charge
sector have been little explored so far. 
 
A class of systems where one can directly investigate these important questions 
is represented by the organic triangular compounds of the family
$\theta$-(BEDT-TTF)$_2$$MM'$(SCN)$_4$ (in short $\theta$-$MM'$) where $M=$Tl,Rb,
Cs and $M'=$ Co,Zn, which exhibits a notable degree of charge frustration 
\cite{Kagawa2013}. These materials are quarter-filled, i.e. they have one
particle per two lattice sites, and in most cases they display a first order
structural transition upon cooling, leading to a charge-ordered insulating
ground state. Such a structural transition, however, can be avoided by
sufficiently rapid cooling, and the system is found to remain a poor conductor,
displaying characteristic kinetic slowing down
\cite{monceau2007prb,Kagawa2013,Sato2014PRB}  reminiscent of glassy dynamics.
Remarkably, this behavior was reported in high quality single crystals, having no
significant amount of disorder, suggesting \cite{Kagawa2013}  a possibility of
self-generated (disorder-free) glassy behavior of electrons.

Despite these significant experimental advances, the following
questions remain: (1) What is the dominant physical mechanism that may produce
such glassy dynamics of electrons? (2) What is the role of geometric frustration
and the range of electron-electron interactions? In this Letter we present a
model calculation that provides a clear and physically transparent answer to
these important questions. We show that the long-range nature of the Coulomb
repulsion plays a crucial role in lifting the massive ground-state degeneracy
produced by geometric frustration. However, it does so very weakly, producing an
extensive manifold of low-lying metastable states, causing slow relaxation and
glassy dynamics up to temperatures above the stripe melting transition, even in absence of disorder.

\emph{Model.} --- We study a system of spinless electrons  on a triangular lattice
with inter-site repulsion $V_{ij}$, 
\begin{equation} H = -t \sum_{\langle i,j \rangle} c^\dagger_{i} c_j +
\frac{1}{2} \sum_{ij} V_{ij} \left(n_i - \frac{1}{2} \right) \left(n_j -
\frac{1}{2} \right) \label{ModelH} 
\end{equation} 
where $t$ is the hopping
integral between two neighboring sites and $c^\dagger_i$ is the electron
creation operator. We fix the density to one particle per two sites, $n
= 1/2$, as appropriate to the $\theta$-$MM'$ salts. We also
allow for deviations from the perfect triangular lattice, by defining an
anisotropy parameter $\lambda$  via the unit vectors $a = \hat{y}$ and $b = \frac{1}{2}
\left(\sqrt{3} \lambda \hat{x} + \hat{y}\right)$
($\lambda=1.16$ for $\theta$-CsZn and $\lambda=1.26$ for
$\theta$-RbZn.\cite{Mori1998}). In what follows we shall consider a maximally frustrated
isotropic lattice ($\lambda=1$) unless otherwise specified.

\begin{figure} [t] \includegraphics[width=\columnwidth]{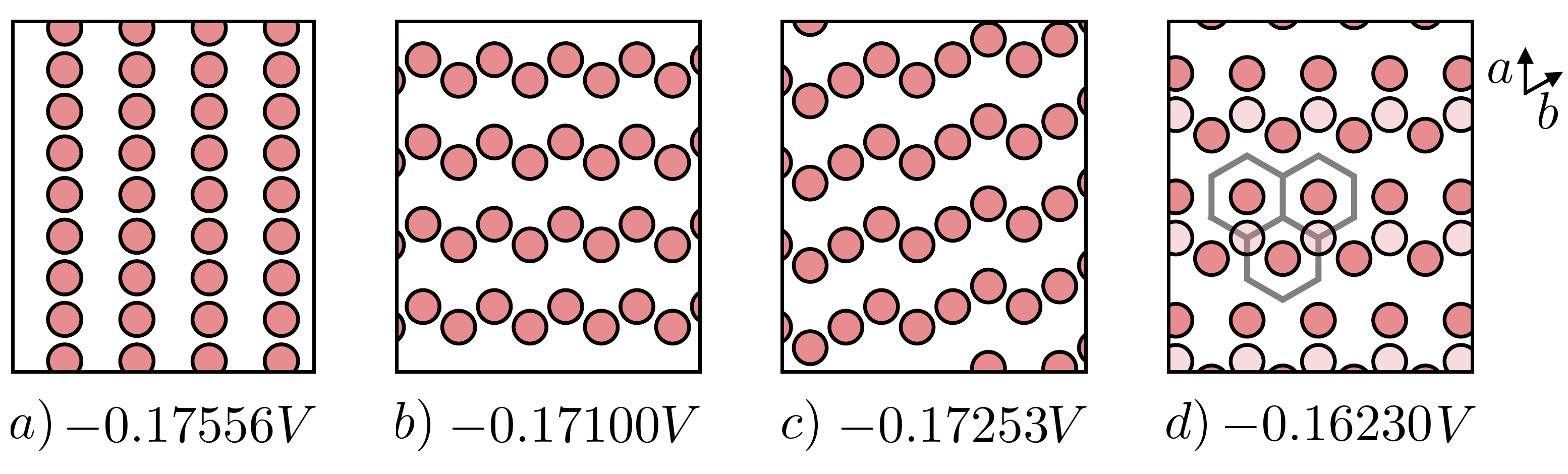}
\caption{Periodic charge configurations on the triangular
lattice with $n=1/2$ particles per site: a) linear stripes, b) short zigzag stripes, c) long zigzag stripes and
d) the lowest energy three-sublattice structure (see text), with the corresponding 
energies calculated for the long range Coulomb potential [$x=1$ in Eq. (\ref{LRI})]. 
For nearest-neighbor interactions only ($x=0$), all these states are
degenerate in energy, $\epsilon=-V/4$. } \label{FigDegenerateStates} \end{figure}

Charge ordering in organic conductors is customarily attributed to the presence of 
strong nearest neighbor repulsion, 
of the form $ V_{ij} = V \delta_{|R_i-R_j|=1}$\cite{Seo}. In this case the ground state of
Eq. (\ref{ModelH}) is massively degenerate in the classical limit $t=0$, as any
state where each particle has exactly two nearest neighbor sites occupied has an
energy per site of $\epsilon=-V/4$. Such ordered states  include the linear (vertical) 
and zigzag (horizontal) 
stripes depicted in Fig.  \ref{FigDegenerateStates}(a) and
(b), as well as any other striped configuration such as the long zigzag stripes of
Fig. \ref{FigDegenerateStates}(c).
%{\tt would the message be clearer if we give a name to this manifold?}

Interestingly, another  class of  configurations with exactly the same potential
energy can be constructed by dividing the lattice into three sublattices, one of
which is filled (``pins"), one empty, and the third is randomly
occupied by the remaining particles (``balls"). 
Fig.  \ref{FigDegenerateStates}(d)
shows one particular example in this class, where the balls, shown as light disks, are 
themselves ordered. 
When quantum effects are turned on, the latter can 
delocalize as Bloch waves on the honeycomb lattice not occupied by the localized pins, 
leading to a unique ground state whose energy is lower than the 
stripes, which is termed \textit{``pinball liquid"}  \cite{HottaFurukawa}. 
This occurs because the delocalization of balls in 
the pinball state yields a kinetic energy gain
$\propto t$, as opposed to stripes, where only local fluctuations $\propto t^2/V$ are allowed. 
For nearest neighbor 
interactions, quantum fluctuations therefore immediately lift the classical degeneracy, as soon as $t\neq 0$.

Real electrons, however, interact through long-range Coulomb interactions. The
central observation of this Letter is the fact that the 
inclusion of an even modest long-range component completely changes the behavior, 
and dominates over the quantum effects described above at least in the
strong coupling regime ($V/t \gg 1$). To make this point, we
consider a family of inter-site interactions of the form
\begin{equation} V_{ij} = (1-x)V \delta_{|R_i-R_j|=1} + x
\frac{V}{|R_i-R_j|}.\label{LRI} 
\end{equation} 
with $x\le 0 \le 1$.
It reduces to the nearest neighbor  interaction for $x=0$, and to the full long range Coulomb
potential for $x=1$.

\emph{Stripe order and metastability.}--- The energies of different charge configurations in the presence of the interaction
Eq. (\ref{LRI}) can be evaluated using standard Ewald summation techniques \cite{Toukmaji}. 
We find that the crucial effect of adding a long-range component is the  lifting of
the ground-state degeneracy already at the classical level, with the linear stripes
 of Fig. \ref{FigDegenerateStates}(a) having a lower 
energy than any other configuration as soon as $x>0$.
The numerical 
results for the full Coulomb potential ($x=1$) are reported in Fig. \ref{FigDegenerateStates} for a set of relevant examples.

First, the resulting finite (albeit small) electrostatic energy gap, $\gtrsim 0.01 V$, separating three-sublattice configurations 
(Fig. \ref{FigDegenerateStates}(d))  from the stripes (Fig. \ref{FigDegenerateStates}(a))
 implies that stripe order must survive as the ground state 
 even in the presence of quantum fluctuations, 
within a \textit{finite}
interval of $t$ in which the quantum processes associated with the pinball
phase can be neglected. 
We confirmed this result  by
performing Exact Diagonalization (ED) calculations
on finite-size clusters of up
to 16 sites, using again Ewald potentials to account for the long range tail of the interaction.
Our ED results show that for the full Coulomb potential ($x=1$) 
a first order transition from stripes to pinball occurs  at $(t/V)_c\simeq 0.05$. 
The stability of stripes is further enhanced 
with increasing anisotropy \cite{HottaFurukawa}, resulting in a critical $(t/V)_c$
which rapidly increases with $\lambda$ (the critical value is 
$(t/V)_c\simeq 0.08$ already at $\lambda=1.16$). Both the rigorous
stability argument given above (for infinitesimal $t$) and our ED results (for
finite $t$) therefore indicate that the behavior obtained from a classical model, where stripes are the lowest energy state,
should be at least qualitatively valid to describe systems with a small ratio $t/V$. This regime
is actually relevant to the organic salts under study, where ab-initio calculations suggest typical values
$t/V\sim 0.05-0.1$ \cite{Nakamura2009,Dolfen}. 
We shall therefore consider in what follows the
$t=0$ limit of Eq. (\ref{ModelH}).
%, and focus in what follows on the effects of  classical fluctuations.

%We start by calculating the energies of different charge configurations in the
%classical limit. 
%Our results show that the qualitatively same behavior is obtained for any $x>0$: as soon as 
%the long-range tail of the interaction potential is turned on, 
%the ground state  is given by the linear stripes. All the other states, which were degenerate in the short range
%model, move to higher energy. 

Second, the manifold of states that were degenerate in the short range model now gives rise to 
a macroscopic number of quasi-degenerate configurations, that are spread 
within typically $0.01V$ from the ground state.
Evaluating the strength  $V = e^2 / \epsilon a$ of the Coulomb potential, with typical values 
$\epsilon \approx 3$ for the dielectric constant and $a \sim 5 \AA$ for the
lattice constant, yields  $V \sim 1 $ eV and $0.01V
\approx 100$ K.  This estimate shows that these configurations are within the thermally accessible range, so that 
they should participate in the classical fluctuations of the electronic system. 
%(relevant examples of these states are illustrated in Fig.
%\ref{FigDegenerateStates}, with the corresponding electrostatic energies for $x=1$).

The reason behind 
%The origin of
such a low energy scale is
that energy differences only depend on the interaction between distant neighbors
(as pointed out above, nearest neighbor interaction terms are the same by construction for all the configurations
in the manifold);
the longer the length scale $|R_i-R_j|$, the more the considered charge configurations will look uniform, and
the interaction energy will eventually average out to very comparable values. 
The barriers between these states are, however, much larger. 
They are actually of order $V$, because going from one configuration to the other necessarily 
involves  local rearrangements of the charge. As we proceed to show, 
such large {\it local} energy scale, contrasted to the much smaller 
{\it global} energy scale corresponding to long-range rearrangements, causes
the system to get easily ``stuck'' in one of
these states, which  therefore become metastable.

\begin{figure}[h] \includegraphics[width=\columnwidth]{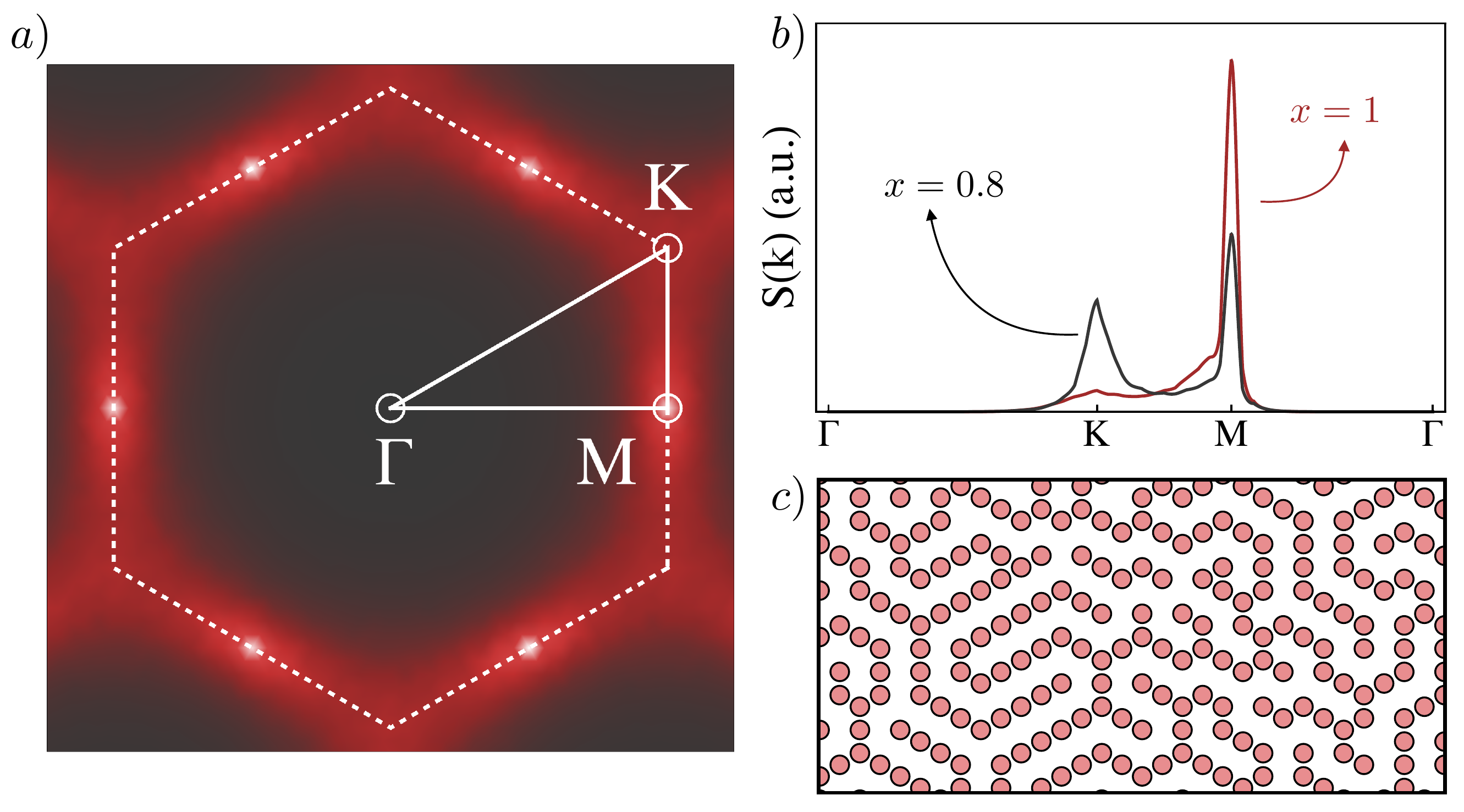}
\caption{
(a)  %Fourier transform of the density-density correlation function 
Structure factor $C(k)$ of the frustrated Coulomb liquid
obtained for a $L=36$ lattice size and averaged over 1000 independent equilibrated Monte Carlo runs at
$T=0.04V$ above the melting transition. Clearly defined diffuse Bragg peaks are observed at the (linear) stripe order
wavevector M $= (\frac{2 \pi}{\sqrt{3}},0)$ and symmetry
equivalent points. (b) Same, plotted along the high-symmetry lines of the Brillouin zone drawn in panel  (a) (arb. units).  
(c) A snapshot illustrating a typical ``stripe liquid" configuration. 
} \label{MetastableFig} \end{figure}

\emph{Glassy behavior.} --- 
To illustrate the dynamic slowing down in the correlated liquid above the freezing transition, reflecting the emergence of many metastable states, we
examine our system at finite temperature through
classical Monte Carlo simulations. We use the Metropolis algorithm with local (nearest neighbor) updates\cite{Grempel} to mimic real-time dynamics of hopping electrons, since in the organic materials the electrons have exponentially suppressed beyond-nearest neighbor hopping. The long-range nature of the interaction is taken into account using Ewald summation\cite{Toukmaji}, with lattice sizes $L=12, 24, 36$ and $48$. Note that $L=48$ corresponds to $L^2/2=1152$ electrons.

Starting from zero temperature with the linear stripe ordered phase, we find that
for the pristine Coulomb potential ($x=1$), the stripe order remains stable up to the  temperature  $T_c
\sim 0.038 V$, where the system undergoes a first-order transition to an
(isotropic) fluid phase with significant short-range order. 
A detailed characterization of the first-order stripe-melting transition is beyond the scope of this work. 
We shall here focus our full attention to understanding the structure and the dynamics of
the correlated fluid phase above the melting transition. 

A typical fluid
configuration is shown in Fig. \ref{MetastableFig}(c), where we observe
finite-size striped domains with random orientations (linear stripes 
are 6-fold degenerate, corresponding to  3 orientations and 2 sublattice origins).
To quantify the observed short-range order, we computed the structure factor
within the correlated fluid phase, defined as the thermally
averaged density-density correlation function $S(\k) = \langle n_\k n_{-\k}
\rangle$. Typical results obtained at
$T=0.04V$, just above the melting transition,  are displayed in Fig. \ref{MetastableFig}. For these results, we equilibrated 1000 independent simulations, with 50,000 measurement sweeps per simulation. 
A sharp peak is found at
the M points, signalling the existence of local stripe order, corresponding to
the linear stripes of Fig. \ref{MetastableFig}(a).  Remarkably, this peak coexists with
a broader feature at the K points, reminiscent of the classical three sublattice 
configurations which are precursor to the pinball liquid phase. 
Note that the relative weight of these two features is directly controlled by 
the long range strength $x$, as shown  in Fig. \ref{MetastableFig}(b).
A diffuse background is also visible along the whole Brillouin zone edge,
indicative of a correlation hole around each electron. The coexistence of two
competing ordering wave-vectors and the emergence of diffuse lines in the
structure factor is a distinctive feature of $\theta$-$MM'$ salts that is
naturally captured when long range interactions are included.

\begin{figure}[t] \includegraphics[width=0.45\textwidth]{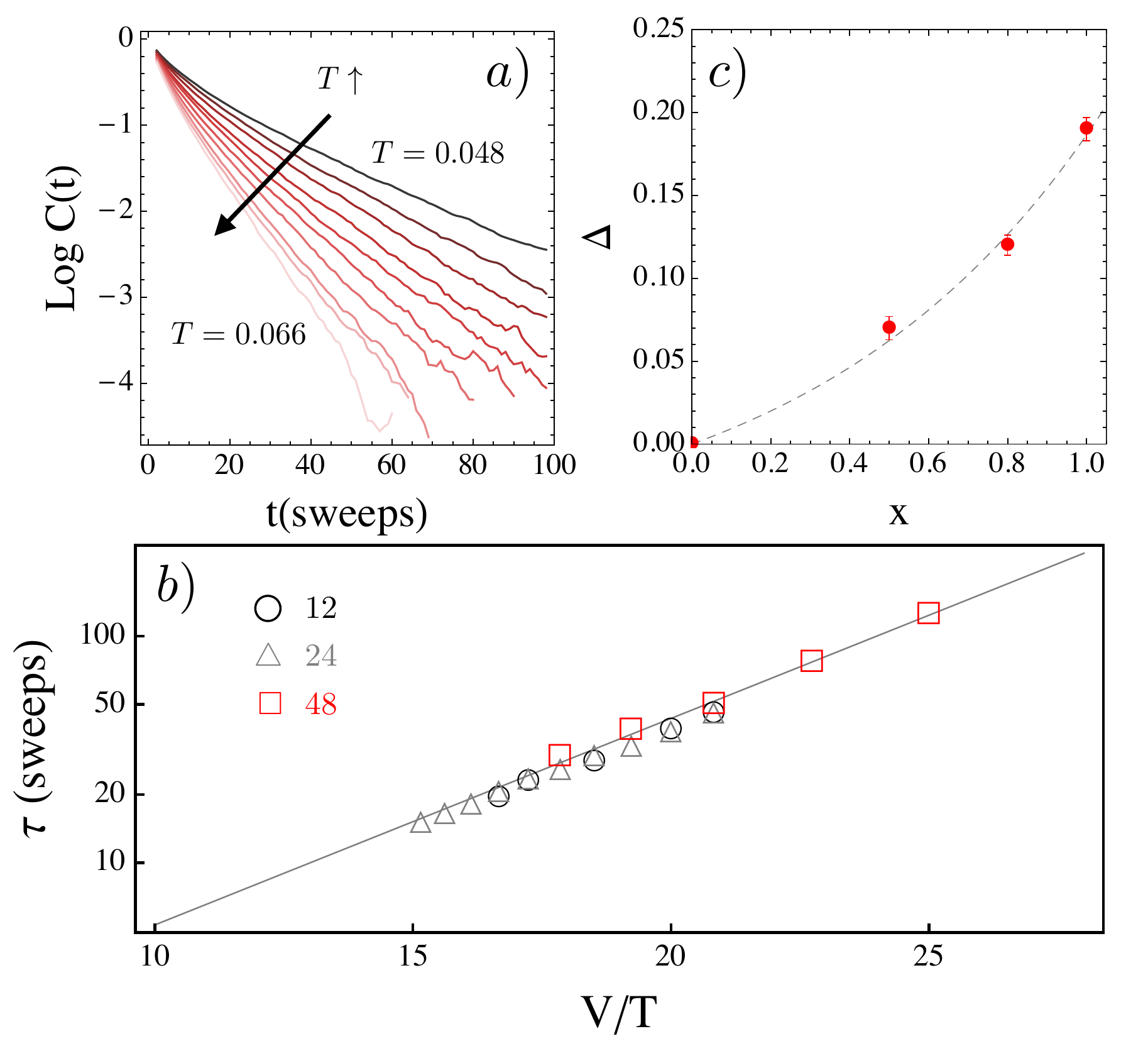}
 \caption{ (a) The local autocorrelation function $C(t)$
of the stripe liquid obtained via Monte Carlo simulations on
lattices  of sizes $L=12, 24, 48$ at different temperatures,
showing exponential decay at long times 
(time in units of the elementary MC step). 
 (b) For all studied sizes, the decay time $\tau$ follows 
a clear Arrhenius scaling $\tau \sim
e^{\Delta/T}$ in the studied interval, signalling strong glassy behavior. (c) The activation energy $\Delta$
as a function of the strength $x$ of the long range interaction tail.}
\label{FigRelaxTime} \vspace{-12pt} \end{figure}

To further characterize the role of striped correlations in the fluid phase, we
turn our attention to the dynamics and set out to describe the dynamic relaxation
processes, closely following the approaches previously used   to
investigate (disordered) Coulomb glasses \cite{Grempel}. Our runs were $5 \times 10^5$ Monte Carlo sweeps long, where one sweep constitutes $L^2$ update attempts. Physical quantities were monitored as a function of time (measured in number of sweeps) for each sample and the results were averaged over between 500 and 1000 initial random configurations.

 We computed the local 
autocorrelation
function
\begin{equation} C(t+t_w,t_w)=\frac{2}{N} \sum_i \langle \delta
n_i(t+t_w) \delta n_i(t_w)\rangle 
\end{equation} where $t_w$ is the waiting time
measured in Monte Carlo sweeps. Following Ref. \cite{Grempel}, we first quickly
``quench" (cool down) the system from a random initial configuration to the desired
temperature $T$, and then allow the system to relax for a waiting time $t=t_w$,
before collecting data for the autocorrelation function. We perform such studies
at each temperature for several values of $t_w$; the system is equilibrated when
 $t_w$ is sufficiently long so that the autocorrelation function becomes
independent of the waiting time, $C(t+t_w,t_w)=C(t)$. The long-time exponential tail of $C(t)$ [Fig.\ref{FigRelaxTime}(a)] then
directly gives us the desired relaxation time $\tau(T)$, which we generally find
to be comparable to the required time to equilibrate the system. The relaxation
time thus found displays Arrhenius behavior \cite{Debenedetti2001},
\begin{equation} \tau (T) = \tau_0 e^{\Delta /T} \label{eq:tau}
\end{equation} 
as illustrated in Fig. \ref{FigRelaxTime}(b) for $x=1$.
The strong (that is, Arrhenius instead of the 'fragile' Vogel-Tammann-Fulcher) glass behavior obtained here is reminiscent of the precursor of the cluster glass phase in
$\theta$-CsZn, as shown by the resistivity aging and noise experiments performed
in Ref. \cite{Sato2014PRB}, where it was found that $\Delta \approx 2600$ K. 
Setting  $V \sim 1 $ eV, the activation energy $\Delta \approx 0.2V$ found in the 
simulations for $x=1$ [Fig. \ref{FigRelaxTime}(c)] corresponds to $2300$K.
This result is of the same order of magnitude as the experimental 
value, providing  quantitative support to the present theoretical picture. 

The dramatic slowing down arising in this correlated stripe liquid directly
reflects the long-range nature of the Coulomb interaction. To illustrate this
point, we gradually decrease the amplitude $x$ of the long-range Coulomb
potential. We find that the same qualitative behavior is found for any non-zero value of
$x$, and the relaxation time still displays simple activated behavior, however
with an $x$-dependent activation energy, which is seen to decrease as $x$ is
reduced, and vanish at $x=0$ [Fig. \ref{FigRelaxTime}(c)]. 
\cite{mori2003non}

\begin{figure}[t] \includegraphics[width=\columnwidth]{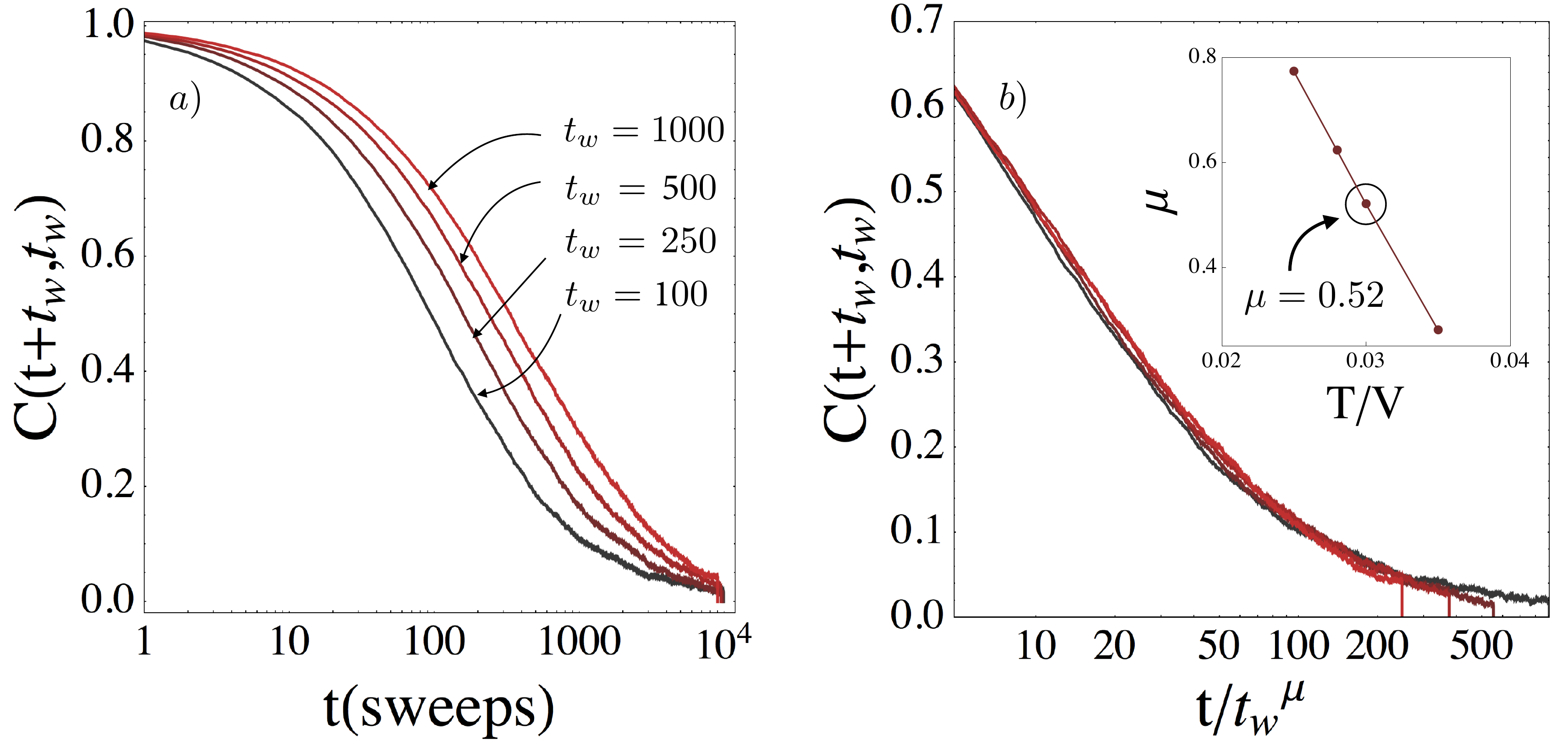}
\caption{ 
(a) Autocorrelation function for different waiting times, at $T=0.03$. Equilibrium is
restored only if the waiting time $t_w$ is sufficiently long; the
corresponding relaxation time $\tau \approx 10^3$, and for $t_w < \tau$, the
system displays history dependence and aging, as typically found in glass
formers in the supercooled liquid regime. (b) 
Data can be collapsed in terms of $t/t_w^\mu$, with $\mu$ a temperature dependent 
aging exponent  (inset), similarly as in (disordered) Coulomb
glasses \cite{Grempel}. } 
\label{FigAging}\vspace{-18pt} 
\end{figure}

The results presented above correspond to an equilibrated fluid state at temperatures 
above the crystal melting temperature $T_c$. Experimental results for $\theta-MM'$ 
compounds indeed show that the dynamical slowing down and the short-range charge 
correlations are already manifest above the melting temperature, as a precursor 
for further glassy freezing in the supercooled regime \cite{Kagawa2013,Sato2014PRB}.  
At temperatures below the stripe melting transition, the relaxation time becomes very
 long, which allows a study of the dynamics for waiting times shorter than $t_w <
\tau(T)$. In this regime the autocorrelation
function depends on both $t$ and the waiting time $t_w$, displaying
characteristic ``aging" behavior \cite{Grempel}. Here, the autocorrelation
function $C(t+t_w,t_w)$ assumes a scaling form $F(t/t_w^\mu)$ where $\mu$ is the
aging exponent, as illustrated in Fig.~\ref{FigAging}. We found such aging
behavior in the entire supercooled regime $T < T_c$, demonstrating dynamical behavior
precisely of the form expected for supercooled liquids around the glass
transition, consistent with results obtained for Coulomb glasses 
\cite{Grempel}. In fact, our qualitative and even quantitative results are very
similar to this well-known glass former, indicating that robust
glassy behavior emerges in our model even in the absence of disorder.

\emph{Outlook.} --- We have shown that the interplay of long range interactions and 
geometric frustration plays a singular role in
Coulomb liquids, producing a multitude of metastable states, slow relaxation,
and most characteristic features of disorder-free glassy dynamics in the correlated liquid regime, 
as recently observed in the two-dimensional organic conductors
$\theta$-$MM'$. The general mechanism identified here, however, may 
be expected to be significant in a much broader class of systems,
including many families of complex (e.g. spinel) oxides with not only
triangular, but also Kagome and pyrochlore lattices.
Quantum fluctuations can be conveniently tuned
in these systems, for example by applying pressure or appropriate chemical
doping. The resulting quantum critical behavior of such self-generated Coulomb
glasses thus arises as one of the most attractive avenues of experimental and
theoretical  investigation in future work, not only from the perspective of basic science 
research, but also as an important issue for the next generation of electronic devices \cite{tokura2014prbR}.

 The authors thank K. Kanoda, T. Sato, F. Kagawa, L.
Balents, Z. Nussinov and J. Zaanen for interesting discussions. L.R. was supported by the
Dutch Science Foundation (NWO) through a Rubicon grant. This work is supported
by the French National Research Agency through Grant No. ANR-12-JS04-0003-01
SUBRISSYME. Work in Florida (S. M. and V. D.) was   supported by the NSF grants
DMR-1005751 and DMR-1410132. V.D. would like to thank CPTGA for financing a
visit to Grenoble, and KITP at UCSB, where part of the work was performed.

\bibliographystyle{apsrev}

\bibliography{glass}

\end{document}